\documentclass[sigconf,screen]{acmart}

\usepackage{subcaption}

\renewcommand\footnotetextcopyrightpermission[1]{} 

\AtBeginDocument{}

\acmConference[OSCAR 2025]{The 4th Workshop on Open-Source Computer Architecture Research}{June 21, 2025}{Tokyo, Japan}

\begin{document}

\title{Revisiting Hardware Priority Queue Architectures}
\author{Qihang Wu}
\affiliation{
	\institution{New York University}
	\department{Electrical and Computer Engineering}
	\city{Brooklyn}
	\state{NY}
	\country{USA}
}
\email{qw2246@nyu.edu}

\author{Austin Rovinski}
\affiliation{
	\institution{New York University}
	\department{Electrical and Computer Engineering}
	\city{Brooklyn}
	\state{NY}
	\country{USA}
}
\email{rovinski@nyu.edu}

\maketitle

\section{Introduction}

Priority queues — data structures that serve elements based on priority rather than insertion order — are fundamental in a
wide range of applications, including operating systems, graph algorithms, and data compression. Software
implementations, typically based on binary heaps with \(O(log\ N)\) complexity, are sufficient for many scenarios; however they
can become performance bottlenecks in latency-sensitive domains such as networking and robotics. Hardware-based priority queues exploit parallelism to significantly reduce operation latency, delivering critical performance improvements in latency-sensitive applications.

Despite the breadth of prior work on hardware priority queues, two major challenges remain. First, many foundational
architectures were proposed and studied years ago, calling into question their relevance given modern hardware advancements.
Second, comprehensive comparisons across different architectures are lacking, making it difficult to evaluate
trade-offs in performance, resource utilization, and scalability. This paper addresses both gaps by implementing and
evaluating several representative hardware priority queue architectures on modern FPGA platforms and providing a
quantitative analysis to guide future design choices. All implementations, tests, and analyses are available through our
open-source library at \url{https://github.com/realise-lab/hwpq}.

\section{Overview}

\textbf{Previous Works}. Over the past two decades, several significant contributions have advanced the design of hardware priority queues. Moon et al.~\cite{moon2000} evaluated existing architectures and introduced both a modified systolic array and a
multiple-output-link priority queue, achieving constant-time operations to address scalability challenges in network
switches. Huang et al.~\cite{huang2014} focused on database applications, proposing three FPGA-based designs: a
register-based array, a register-based tree, and a BRAM-based tree, culminating in a hybrid tree architecture that
combined register arrays with multiple BRAM-based trees for enhanced performance. Zhou et al.~\cite{zhou2020} applied
hardware priority queues to real-time path planning, developing a systolic array-based architecture with an efficient
cache system to accelerate the A* algorithm. Collectively, these contributions demonstrate how hardware-specific optimizations 
can significantly enhance performance across various domains by leveraging custom circuit designs and parallelism.


\textbf{Architecture Library.} Currently implemented architectures are listed in Table~\ref{tab:arch-comparison}. Some implementation have the ability to disable support for enqueue operation in order to improve efficiency for dequeue and replace operations.

\begin{table}[h]
  \small 
  \centering
  \caption{Priority Queue Architectures in Library}
  \label{tab:arch-comparison}
  \begin{tabular}{@{}lccc@{}}
    \toprule
    \textbf{Architecture} & \textbf{Enq. Switch} & \textbf{Enq. Latency} & \textbf{Rep./Deq. Latency} \\
    \midrule
    Reg. Array           & Yes & 1          & 1          \\
    Reg. Array Pipe.     & Yes & 2          & 2          \\
    Reg. Tree            & Yes & $\log N$   & 1          \\
    Reg. Tree Pipe.      & Yes & $\log N$   & 2          \\
    Systolic Array       & Yes & 1          & 1          \\
    BRAM Tree            & No  & N/A        & 8          \\
    BRAM Tree Pipe.      & No  & N/A        & 4          \\
    Hybrid Tree          & No  & N/A        & 1          \\
    \bottomrule
  \end{tabular}
\end{table}

\textbf{Methodology.}
%
%
%
To ensure a uniform evaluation across all priority queue architectures, a standardized interface was developed. Each
design was synthesized and implemented using AMD Vivado targeting the Artix UltraScale+ FPGA (XCAU25P). A comprehensive parameter sweep was conducted, varying queue size, enqueue support, and pipelining strategies. For each configuration, we measured the maximum operating frequency and resource utilization, including lookup tables (LUTs),
flip-flops (FFs), and block RAMs (BRAMs). This methodology allowed us to assess both performance and resource efficiency while minimizing confounding effects from platform limitations. The FPGA platform ensured that observed characteristics reflected inherent architectural trade-offs rather than implementation constraints. Detailed implementation information and FPGA synthesis results are available in our repository~\cite{hwpq-repo}.

\textbf{GitHub Repository and Usage.} Our open-source hardware priority queue library provides modular,
parameterized RTL implementations with a standardized interface that allows designers to easily swap between different
architectures without modifying the surrounding logic. Each implementation includes configurable parameters such as
queue size, data width, and an enable switch to turn on/off support for enqueue operations (register-based architectures only). Comprehensive documentation, verification testbenches, and synthesis scripts are provided to streamline integration and evaluation. Given the complex
performance–resource trade-offs associated with different architectures and workloads, the library's modular design
enables direct integrations and comparisons based on specific application requirements, allowing designers to identify optimal solutions without relying on generalized benchmarks.

\vspace{-2mm}
\section{Results and Takeaways}
In this section, we present some high-level takeaways from our detailed exploration of architectures in our library. Our GitHub repository features a more complete and detailed comparison of the architectures~\cite{hwpq-repo}.


\begin{figure*}[h]
    \centering
    \begin{subfigure}[h]{0.49\textwidth}
	   \centering
	   \includegraphics[width=0.7\linewidth]{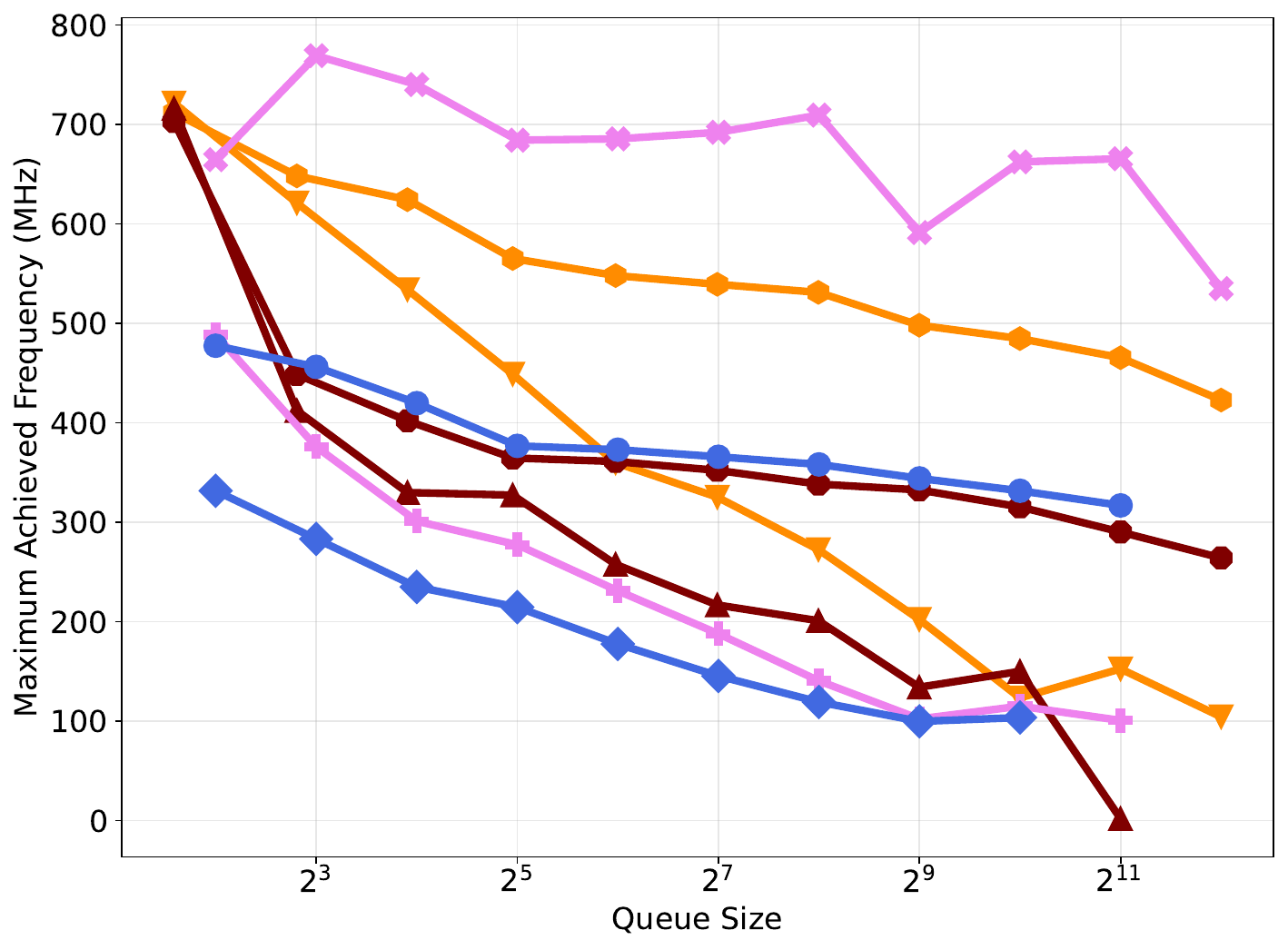}
	   \caption{Register-based Architectures' Achieved Frequency}
	   \Description{Register-based architectures' achieved frequency cross comparison}
	   \label{fig:reg-based-freq}
    \end{subfigure}
    \hfill
    \begin{subfigure}[h]{0.49\textwidth}
	   \centering
	   \includegraphics[width=0.7\linewidth]{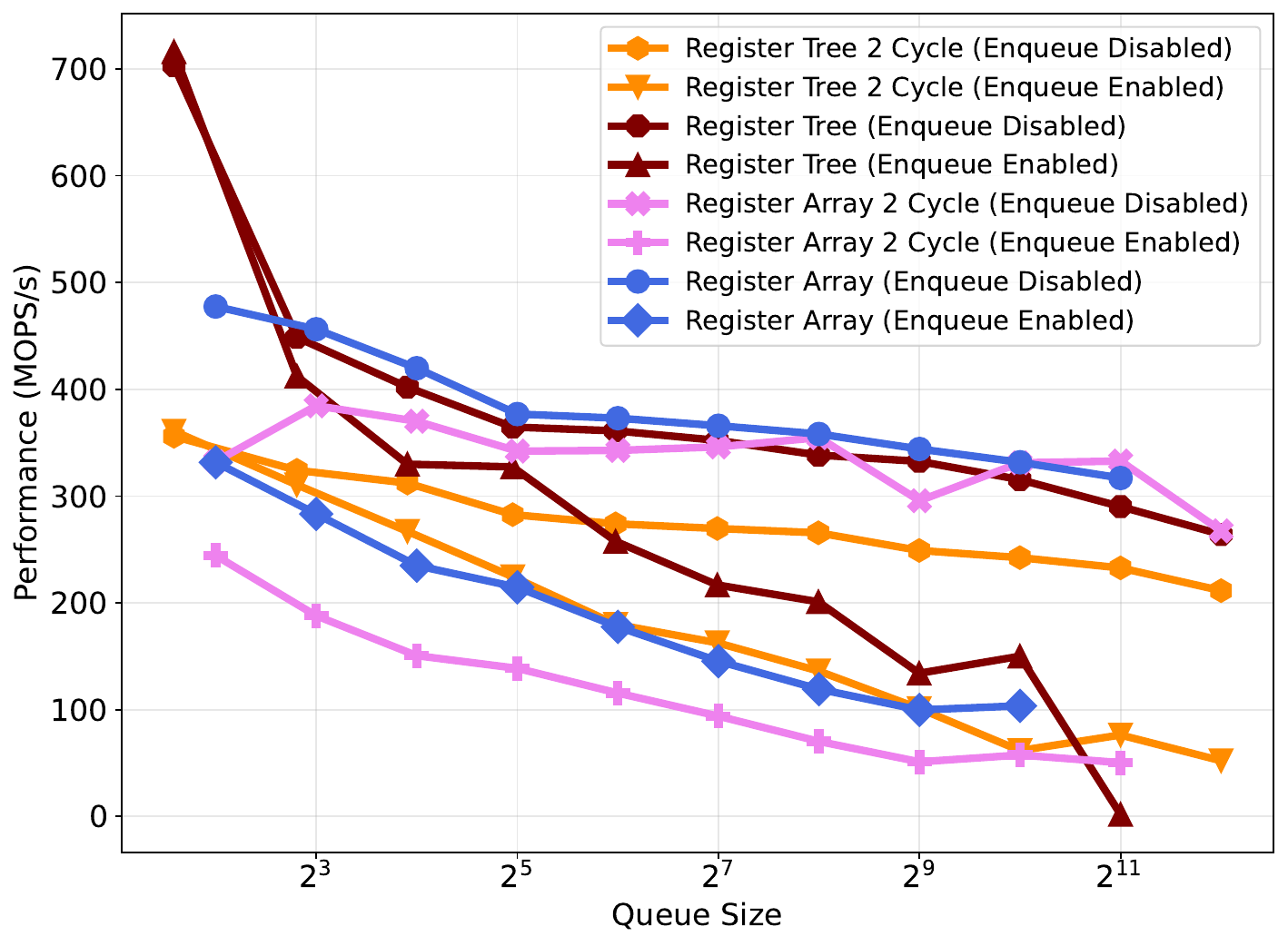}
	   \caption{Register-based Architectures' Performance}
	   \Description{Register-based architectures' performance cross comparison}
	   \label{fig:reg-based-perf}
    \end{subfigure}
    \caption{Register-based Architectures Achieved Frequency and Performance}
    \label{fig:reg-arch-freq-and-perf}
\end{figure*}

\begin{figure*}[h]
    \centering
    \begin{subfigure}[h]{0.49\textwidth}
	   \centering
	   \includegraphics[width=0.7\linewidth]{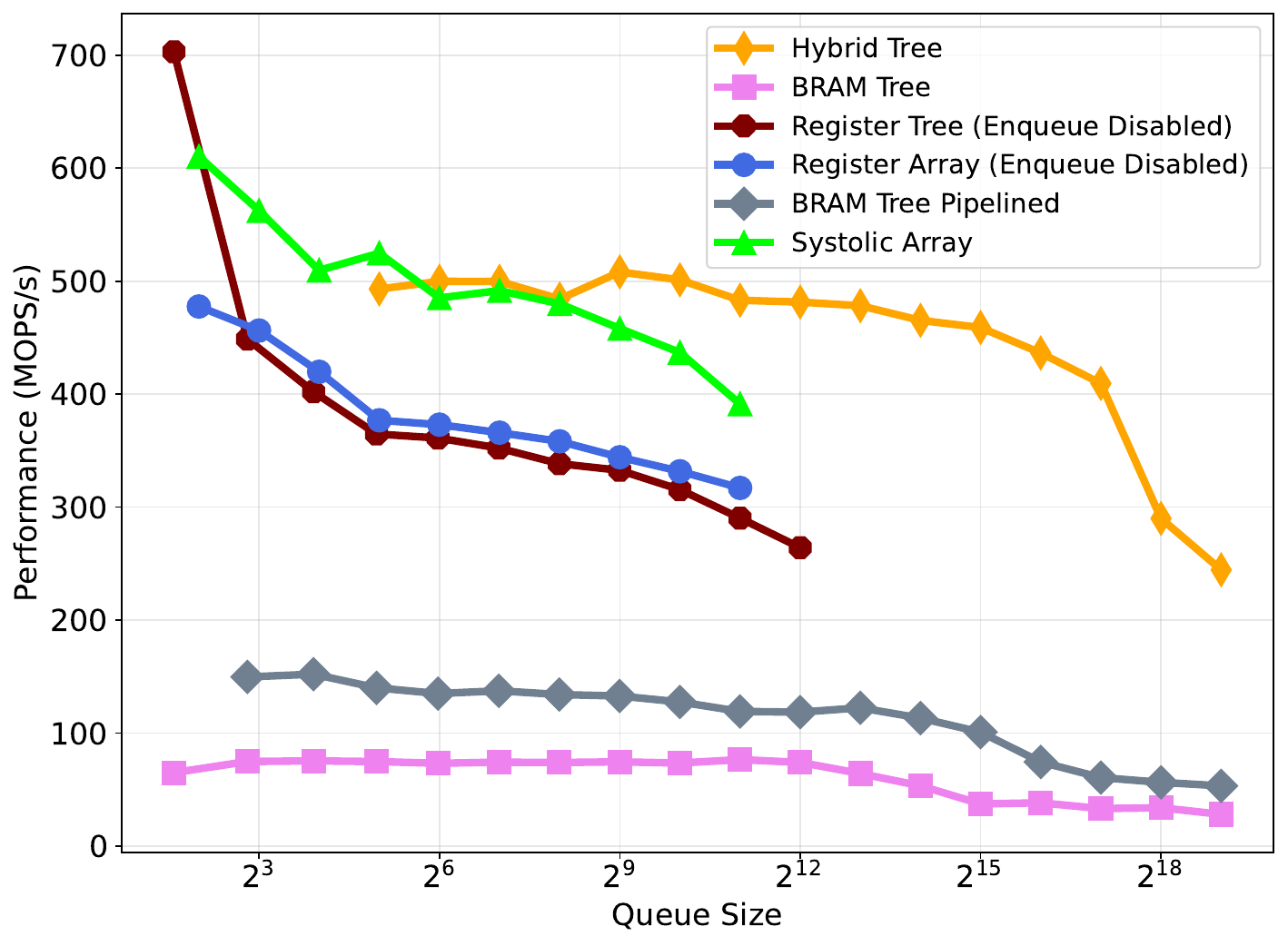}
	   \caption{Selected Architectures' Performance}
	   \Description{Overall architectures performance comparison}
	   \label{fig:arch-perf-comp}
    \end{subfigure}%
    \hfill
    \begin{subfigure}[h]{0.49\textwidth}
	   \centering
	   \includegraphics[width=0.7\linewidth]{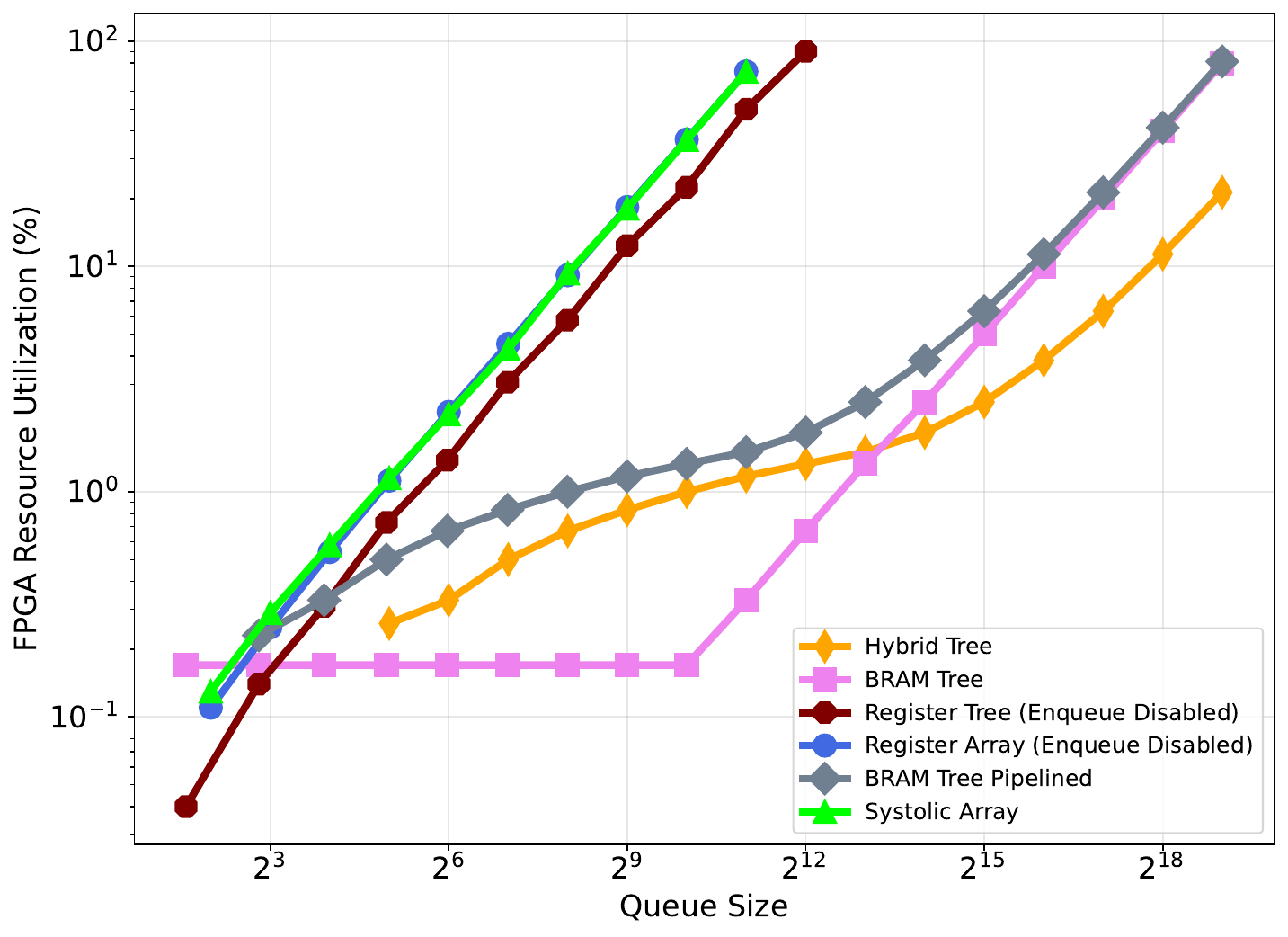}
	   \caption{Selected Architectures Resource Utilization}
	   \Description{Overall architectures' resource utilization comparison}
	   \label{fig:arch-resource-comp}
    \end{subfigure}
    \caption{Selected Architectures Performance and Resource Utilization Comparison\vspace{-3mm}}
    \label{fig:arch-perf-and-resource-util}
\end{figure*}

\textbf{Enqueue vs. Non-enqueue Support.}
%
%
The performance impact of supporting dynamic enqueue operations versus optimizing solely for replace operations, as
explored by Huang et al.~\cite{huang2014}, was evaluated. Figure~\ref{fig:reg-based-perf} shows that enabling enqueue 
support incurs significant performance overhead, particularly for register arrays, with a consistent performance gap 
across all queue sizes. In contrast, for register trees, the performance penalty is minimal for small
queues but increases with larger queue sizes. These results suggest that designers should carefully evaluate whether the
flexibility of dynamic enqueues justifies the associated performance costs based on the specific application
requirements.

%

\textbf{Pipelined vs. Non-Pipelined Implementations.}
%
We explored pipelining the compare-and-swap logic across two clock cycles. While this approach increased the maximum
clock frequency, as shown in Figure~\ref{fig:reg-based-freq}, we found it tends to reduced overall performance, as shown in Figure~\ref{fig:reg-based-perf}. Our analysis shows that for priority queue architectures with relatively simple compare operations, single-cycle designs at lower frequencies consistently outperform pipelined, multi-cycle counterparts.

\textbf{Register-Based vs. BRAM-Based Implementations.} 
%
Register-based designs leverage flip-flops for storage, offering superior access speeds but facing scalability limitations. In contrast, BRAM-based designs utilize dedicated memory blocks, enabling greater storage capacity at the cost of increased access latency. As shown in Figure~\ref{fig:arch-perf-comp}, register-based architectures generally outperform BRAM-based counterparts, with the exception of the hybrid tree. BRAM-based implementations, however, demonstrate more consistent scalability, maintaining reasonable throughput even for larger queues, albeit with lower peak performance. These distinct characteristics define clear operational domains for each approach, with a notable performance crossover observed at approximately 256 entries when comparing the systolic array (the best register-based design) to the hybrid tree (the best BRAM-based implementation).

\textbf{Takeaways and Design Considerations.}
The hybrid tree~\cite{huang2014} provides the best balance across queue sizes by combining register-based elements with BRAM structures, maintaining consistent performance as queue sizes vary. For maximum performance where ample resources are available, systolic arrays~\cite{zhou2020} excel at small to medium queue sizes despite their relatively high resource consumption (Figure~\ref{fig:arch-resource-comp}), while register trees offer optimal efficiency for queues with fewer than 8 entries in resource-constrained applications.


Our results show that the systolic array and hybrid tree architectures offer the best performance. For queue sizes below 64 ($2^6$) entries, the systolic array performs better, as shown in Figure~\ref{fig:arch-perf-comp}. However, beyond this point, the hybrid tree achieves comparable performance while also being more resource-efficient due to its use of both registers and BRAM, as demonstrated in Figure~\ref{fig:arch-resource-comp}. As queue sizes exceed 256 ($2^8$) entries, the hybrid tree surpasses the systolic array in both performance and area efficiency, making it the preferred choice for larger designs.

The main takeaway is that designers will often need to consider their specific use case and consider what power/performance/resource tradeoffs they want to make for their design. We provide the HWPQ library~\cite{hwpq-repo}, which enables designers to configure and plug in a variety of architectures in order to quickly determine the best architecture for their needs.

\bibliographystyle{ACM-Reference-Format}
\bibliography{ref}

\end{document}